\documentclass[useAMS,usenatbib]{mn2e}
\usepackage{graphicx, txfonts, natbib}
\usepackage{epsf}
\usepackage{amssymb}
\usepackage{epstopdf}
\usepackage{longtable}
\usepackage{subfigure}
\usepackage{comment}
\usepackage[colorlinks=false,dvips]{hyperref}
\newcommand{\msun}{\mbox{M$_{\odot}$}}

\newcommand{\kms}{\mbox{$\rm{km}\,s^{-1}$}}

\DeclareMathAlphabet{\mathsc}{OT1}{cmr}{m}{sc}
\def\testbx{bx}%
\DeclareRobustCommand{\ion}[2]{%
\relax\ifmmode
\ifx\testbx\f@series
{\mathbf{#1\,\mathsc{#2}}}\else
{\mathrm{#1\,\mathsc{#2}}}\fi
\else\textup{#1\,{\mdseries\textsc{#2}}}%
\fi}
\newcommand{\ha} {\mbox{H$\alpha$}}

\newcommand{\Nai}{\ion{Na}{i}}

\newcommand{\FeiiF} {[\ion{Fe}{ii}]}

\newcommand{\CaiiF} {[\ion{Ca}{ii}]}

\newcommand{\Hii} {\ion{H}{ii}}
\newcommand{\Hei} {\ion{He}{i}}

\newcommand{\Oi} {[\ion{O}{i}]}

\newcommand{\Oii} {[\ion{O}{ii}]}

\begin{document}
\title[Detecting H and He in late-time SN Ia spectra] {Searching for swept-up hydrogen and helium in the late-time spectra of 11 nearby Type Ia supernovae}
 \author[K. Maguire et al.]
 {K.~Maguire,$^{1,2}$\thanks{E-mail: kate.maguire@qub.ac.uk} S.~Taubenberger,$^{1,3}$ M.~Sullivan,$^4$ P.~A.~Mazzali$^{3,5}$ \\ 
      $^1$European Organisation for Astronomical Research in the Southern Hemisphere (ESO), Karl-Schwarzschild-Str. 2, 85748 Garching b. M\"unchen, Germany\\
      $^2$Astrophysics Research Centre, School of Mathematics and Physics, Queens University Belfast, Belfast BT7 1NN, UK\\
      $^3$Max-Planck Institut f\"ur Astrophysik, Karl-Schwarzschild-Str. 1, 85748 Garching b. M\"unchen, Germany\\
      $^4$Physics \& Astronomy, University of Southampton, Southampton, Hampshire, SO17 1BJ, UK\\
     $^5$Astrophysics Research Institute, Liverpool John Moores University, IC2, Liverpool Science Park, 146 Brownlow Hill, Liverpool L3 5RF, UK\\
      }
\maketitle

\begin{abstract}
The direct detection of a stellar system that explodes as a Type Ia supernova (SN Ia) has not yet been successful. Various indirect methods have been used to investigate SN Ia progenitor systems but none have produced conclusive results. A prediction of single-degenerate models is that H- (or He-) rich material from the envelope of the companion star should be swept up by the SN ejecta in the explosion. Seven SNe Ia have been analysed to date looking for signs of H-rich material in their late-time spectra and none were detected. We present results from new late-time spectra of 11 SNe Ia obtained at the Very Large Telescope using XShooter and FORS2. We present the tentative detection  of \ha\ emission for SN 2013ct, corresponding to $\sim$0.007 \msun\ of stripped/ablated companion star material (under the assumptions of the spectral modelling). This mass is significantly lower than expected for single-degenerate scenarios, suggesting that $>$0.1 \msun\ of H-rich is present but not observed. We do not detect \ha\ emission in the other 10 SNe Ia. This brings the total sample of normal SNe Ia with non-detections ($<$0.001--0.058 \msun) of H-rich material to 17 events. The simplest explanation for these non-detections is that these objects did not result from the explosion of a CO white dwarf accreting matter from a H-rich companion star via Roche lobe overflow or symbiotic channels. However, further spectral modelling is needed to confirm this. We also find no evidence of He-emission features, but models with He-rich companion stars are not available to place mass limits.
\end{abstract}

\begin{keywords}
line: profiles -- supernovae: general
\end{keywords}

\section{Introduction} \label{intro}
The stellar systems and explosion mechanisms that produce Type Ia supernovae (SNe Ia) are still under debate. There are two main competing scenarios: the double-degenerate scenario \cite[DD;][]{1984ApJS...54..335I,1984ApJ...277..355W} where the companion star to the exploding white dwarf is another white dwarf, and the single-degenerate scenario \cite[SD;][]{1973ApJ...186.1007W} where the companion star is a non-degenerate star such as a main-sequence (MS), giant or sub-giant. The `double-detonation' explosion mechanism has also seen a recent resurgence in popularity. In this scenario, the explosion results from the detonation of He on the white dwarf surface (accreted from either a degenerate or non-degenerate companion star), which triggers a subsequent C detonation in the core that unbinds the star \citep{1982ApJ...253..798N,1990ApJ...361..244L,2007ApJ...662L..95B,2014ApJ...785...61S}.

Recent work has highlighted that it is likely that more than one of these channels contributes to produce `normal' SNe Ia \citep{2010Natur.463..924G,2011Natur.480..348L,2011Natur.480..344N,2011Sci...333..856S,2012Sci...337..942D,2012Natur.481..164S,2013MNRAS.436..222M,2013Sci...340..170W,2014ARA&A..52..107M,2014MNRAS.445.2535S}. However, clear observational diagnostics of the progenitor channel for individual objects have still not been determined. 

The presence of circumstellar material (CSM) around SNe Ia was originally suggested as a key observational diagnostic of the SD scenario (expected not to be present in DD scenarios). CSM signatures have been detected in a number of SNe Ia through studies of time-varying \Nai\ D absorption lines in maximum-light spectra \citep{2007Sci...317..924P,2009ApJ...693..207B,2009ApJ...702.1157S}. The presence of CSM has also been inferred through statistical studies that have shown that $\sim$20 per cent of SNe Ia show signs of CSM \citep{2011Sci...333..856S,2013MNRAS.436..222M} and its presence is more common in more luminous SNe Ia \citep{2013MNRAS.436..222M}. However, recent results have also shown at least qualitatively that the observed \Nai\ D absorption profiles could also be produced in merger scenarios and/or where the companion star is a second white dwarf \citep{2013ApJ...772....1R,2013ApJ...770L..35S,2013MNRAS.431.1541S}.

A unusual class of thermonuclear SNe, dubbed `SNe Ia-CSM' \citep{2013ApJS..207....3S}, display strong signatures of H (\ha\ emission with typical widths of $\sim$2000 \kms)  in their spectra at early times, consistent with interaction with a H-rich CSM: SN\,2002ic \citep{2003Natur.424..651H,2004ApJ...605L..37D,2004MNRAS.354L..13K,2004ApJ...604L..53W,2004ApJ...616..339W}, SN\,2005gj
\citep{2006ApJ...650..510A,2007arXiv0706.4088P}, SN\,2008J \citep{2011ApJ...741....7F,2012A&A...545L...7T}, and
PTF11kx \citep{2012Sci...337..942D}. A sample including new objects discovered by the Palomar Transient Factory is discussed in \cite{2013ApJS..207....3S}. 

SNe Ia-CSM are suggested to result from a progenitor channel involving a non-degenerate companion star such as a symbiotic system \citep[e.g.][]{2003Natur.424..651H,2012Sci...337..942D}. Some of these objects, such as PTF11kx, have been observed at late times ($>$200 d post maximum), where their H emission due to interaction with CSM is found to be still strong \citep{2013ApJS..207....3S,2013ApJ...772..125S}. Since these SNe appear very unusual also at early times and perhaps represent a distinct class of SNe Ia, the astrophysical community does not consider them to be `normal' SNe Ia, i.e. SNe Ia that would be used in cosmological studies. 

Another observational signature of a non-degenerate companion star would be shock interaction from the SN ejecta impacting on the companion star, which was calculated by \cite{2010ApJ...708.1025K}. This interaction would take the form of an increased luminosity in the very early-time (a few days after explosion) light curves of SNe Ia but would be dependent on the SN viewing angle. Signatures of this were searched for in a number of optical-wavelength studies but no clear detection was made \citep{2010ApJ...722.1691H,2011ApJ...741...20B,2011MNRAS.416.2607G,2011ApSS.335..223T,2012ApJ...749...18B,2013ApJ...778L..15Z,2015Natur.521..332O}. Recent early photometric observations of a SN Ia, iPTF14atg, have shown evidence of ultra-violet emission soon after explosion, consistent with the predictions of interaction with a non-degenerate companion star \citep{2015Natur.521..328C}. However,  iPTF14atg was not a normal SN Ia; it displayed unusual photometric and spectroscopic properties being most similar to the peculiar class of sub-luminous `SN 2002es-like' SNe Ia \citep{2012ApJ...751..142G}

Given these inconclusive results on the progenitor configuration of `normal' SNe Ia from CSM and shock-interaction studies to date, complementary tracers of progenitor scenarios are very important. One particular area of interest is the detection of hydrogen in late-time ($\gtrsim$200 d) SN Ia spectra -- H features are expected to be present only in the SD scenario, where H-rich material is removed from a non-degenerate companion star. A key benefit of this nebular-phase probe is that the SN ejecta is expected to be optically thin at these late phases, and therefore the detection of H-rich material does not depend on orientation angle. This H emission is distinct from that seen in SNe Ia-CSM at late times, where the broader H emission is due to interaction with H-rich CSM. 

The ablation (heating) or stripping (momentum transfer) of H-rich material from a non-degenerate companion star through the impact of the SN Ia ejecta on the companion was discussed by \citet[]{1975ApJ...200..145W} and \citet{1986SvA....30..563C}. A number of early numerical studies were performed looking at the interaction between the SN ejecta and a non-degenerate companion star \citep{1981ApJ...243..994F,1984ApJ...279..166T,1992ApJ...399..665L}.  The predicted properties of the H- and He-rich material have since been constrained in greater detail in a number of independent studies \citep{2000ApJS..128..615M,2007PASJ...59..835M,2008A&A...489..943P,2010ApJ...715...78P,2012ApJ...750..151P,2012A&A...548A...2L,2013ApJ...774...37L}. 

The amount of unbound material present in the system after explosion, as well as its velocity distribution, is found to depend on the properties of the companion star. Assuming the case of Roche lobe overflow (RLOF), where the binary separation is typically $\sim$3R$_*$ (where R$_*$ is the radius of the companion star), the latest three-dimensional simulations of MS, red giant (RG) and He-star companions find unbound masses of $\sim$0.1--0.2, 0.6 and 0.02--0.06 \msun, respectively \citep{2012A&A...548A...2L,2013ApJ...774...37L,2012ApJ...750..151P}.  The smallest amount of material is removed in the WD+He star scenario since although the initial binary separation is small (a key parameter determining the amount of mass removed through stripping and ablation), a He star is more compact and has a higher binding energy than an MS or RG companion. This makes the removal of its envelope in the explosion more difficult. In the case of the WD+He star scenario, the material is expected to be He- instead of H-rich.  The peak velocities of the unbound material are also predicted to depend on the companion-star properties. In the simulations of \cite{2010ApJ...715...78P,2012ApJ...750..151P}, the peak velocities of the unbound material are $\sim$550, 660, and 955  \kms\ for WD+MS, WD+RG and WD+He star systems, respectively. The \cite{2012A&A...548A...2L,2013ApJ...774...37L} velocity predictions are in good agreement with these values.   

Alternative mass-transfer mechanisms such as WD+RG systems that transfer mass to the white dwarf through a stellar wind (symbiotic systems) can have binary separations larger than those permitted for RLOF systems. Some recurrent nova systems such as RS Ophiuchi and T Coronae Borealis can have separations of up to $\sim$5R$_*$ \citep{1987rsop.book.....B},  greater than the allowed separation for RLOF. However, $>$0.5 \msun\ of material is still expected to be removed from the companion star in this scenario \citep{2012ApJ...750..151P}. 

\begin{table*}
  \caption{Spectral observations of the SNe Ia in our sample.}
 \label{tab:spec_info}
\begin{tabular}{@{}lccccccccccccccccccccccccccccc}
  \hline
  \hline
SN Name &Observation$^a$&Phase$^b$&Date$^b$& MJD$^b$&Instrument&Grism&Exposure$^c$&References for  \\
&MJD&scaled to (d)& scaled to&scaled to&&&time (s)&time of max.\\
\hline
\hline
SN 2009ig&55485.3&405&20101016&55485.3&FORS2&300V, 300I&2400, 2400&\cite{2012ApJ...744...38F}\\
SN 2010gp&55681.4, 55682.3, 55683.4&277&20110501&55682&FORS2&300V, 300I &8100, 8100& \cite{2013AA...554A.127M}\\  
SN 2011ek&56210.2, 56211.2&421&20121010&56210&FORS2&300V, 300I&5400, 5400&\cite{2012MNRAS.426.2359M}\\ 
SN 2011iv&56210.4, 56212.3, 56223.3&318&20121023&56223&FORS2&300V, 300I&1800, 900&\cite{2012ApJ...753L...5F}\\
SN 2012cg&56420.2&339&20130508&56420.2&XSH&--&4800&\cite{2012ApJ...756L...7S}\\
SN 2012cg&56423.2, 56424.1&343&20130512&56424&FORS2&300V, 300I&600, 600&\cite{2012ApJ...756L...7S}\\
SN 2012cu&56424.2, 56445.1&344&20130602&56445&FORS2&300V, 300I&3600, 3600&\cite{2012CBET.3146....2M}\\
SN 2012fr&56600.3&357&20131104&56600.3&XSH&--&6000&\cite{2013ApJ...770...29C}\\
SN 2012ht&56718.2, 56741.2, 56742.2&433&20140312&56728&XSH&--&12600&\cite{2013MNRAS.436..222M}\\ 
SN 2013aa&56696.4, 56714.4&360&20140216&56704&XSH&--&3800&\cite{2013MNRAS.436..222M}\\ 
SN 2013cs&56741.3, 56742.3&303&20140325&56741&XSH&--&11732&\cite{2013CBET.3533....1A}\\
SN 2013ct&56645.1&229&20131219&56645.1&XSH&--&1900&\cite{2013CBET.3539....1P}\\
\hline
\end{tabular}
 \begin{flushleft}
 $^a$MJD = Modified Julian date. The observation MJD refers to the mid-point of each observation block taken for the SN.\\
 $^b$The spectra have been calibrated to match the photometry at this phase, date and MJD.\\
 $^c$Exposure times for FORS2 are for the 300V, 300I grisms, respectively.\\
\end{flushleft}
\end{table*}

Observational studies of seven nearby `normal' SNe Ia have been performed to look for low-velocity ($<$1000 \kms) H-features in their late-time spectra \citep{2005A&A...443..649M,2007ApJ...670.1275L,2013ApJ...762L...5S,2013MNRAS.435..329L,2015A&A...577A..39L}. No positive detections have been made. Three of the late-time spectra studied were of low-resolution ($\sim$700 \kms), comparable to the width of the expected H features: SN 2001el \citep{2005A&A...443..649M}, SN 1998bu, SN 2000cx \citep{2013MNRAS.435..329L}. The spectra used by \cite{2007ApJ...670.1275L} for SN 2005am and SN 2005cf, and by \cite{2013ApJ...762L...5S} for SN 2011fe, were significantly higher resolution ($\sim$150 \kms). A spectrum obtained for SN 2014J, the closest SN Ia for several decades, had a resolution of $\sim$300 \kms  \citep{2015A&A...577A..39L}.

The tightest limits on the H flux present have been obtained for the nearby SN 2011fe \cite[][]{2011ApJ...733..124S}, corresponding to an upper limit of 0.001 \msun\ of swept-up material if a linear extrapolation of the H fraction of the \cite{2005A&A...443..649M} spectral synthesis model holds \citep{2013ApJ...762L...5S}. Possible signatures of H at near-infrared (NIR) wavelengths were investigated for SN 1998bu using the Pa$\alpha$ and Pa$\beta$ lines, but less stringent mass limits of $\sim$0.5 \msun\ were placed \citep{2013MNRAS.435..329L}. 
 
 A search for signatures of a He-rich companion star in the late-time spectra of SNe 2011fe and 2014J was carried out \citep{2015A&A...577A..39L}. Based on the results of \cite{2013MNRAS.435..329L}, it is expected that line fluxes of O and Ca produced in He-rich material would be stronger than those of He. Therefore, the presence of \CaiiF\ and \Oi\ lines with widths of $<$1000 \kms\ was investigated by  \cite{2015A&A...577A..39L} -- they detected no features at the expected wavelengths and placed upper He-mass limits from the \Oi\ 6300 \AA\ line of $<$0.002 and $<$0.005 \msun\ for SNe 2011fe and 2014J, respectively. For SN 2011fe, they suggested that this limit is sufficient to rule out all He-rich companion star models from \cite{2012ApJ...750..151P} 
and \cite{2013ApJ...774...37L}, including for larger-than-RLOF binary separations. 

If the model predictions are correct, then the non-detection mass limits of the literature sample are sufficient to rule out MS or RG companion stars in RLOF or accreting mass via a stellar wind for these seven SNe Ia. However, the sample size is small and it is unclear if these SNe Ia are representative of the full SN Ia population.  Therefore, a study looking at a significantly larger sample of nebular phase spectra of SNe Ia is needed.

In this paper, we present low- and medium-resolution nebular-phase spectroscopic data of 11 nearby SNe Ia obtained using the European Southern Observatory (ESO) Very Large Telescope (VLT) with the XShooter spectrograph \citep{2011A&A...536A.105V} and with the FOcal Reducer and low dispersion Spectrograph \citep[FORS2;][]{1998Msngr..94....1A}.  In Section \ref{obs_data}, we detail the observations and data reduction steps employed. The analysis and search for the predicted H$\alpha$ lines, as well as for Pa$\alpha$ and Pa$\beta$ lines at NIR wavelengths are described in Section \ref{analysis}. In the same section, we also discuss the results of a search for signatures of He-rich material measured using the available \Hei, as well as \CaiiF\ and \Oi, lines. The discussion and conclusions are presented in Section \ref{discussion}. Throughout this paper we assume a Hubble constant, $H_0=70$\,km\,s$^{-1}$\,Mpc$^{-1}$.

\section{Observations and Data Reduction}
\label{obs_data}
 
Nebular-phase spectra ($\gtrsim$230 d) of 11 SNe Ia are used in this study: six obtained using VLT+XShooter and six obtained using VLT+FORS2. One object, SN 2012cg was observed with both instruments and serves as a check of the analysis and methods. We discuss the data reduction for the two instruments separately. 

\subsection{XShooter spectra}
XShooter is an echelle spectrograph with three arms (UVB, VIS and NIR) covering the wavelength range of $\sim$3000--25000 \AA. The spectral format of XShooter is fixed but the resolution can be adjusted using different slit widths. For our data, we used slit widths of 0.8, 0.9 and 0.9 arcsec in the UVB, VIS and NIR arms, corresponding to resolutions of $R$ $\sim$ 6200, 8800 and 5300, respectively. Details of the spectral observations are given in Table \ref{tab:spec_info}.

The spectra were reduced using the REFLEX pipeline (version 2.6) with the XShooter module (version 2.5.0), producing flux-calibrated one-dimensional spectra in each of the three arms \citep{2010SPIE.7737E..56M,2013A&A...559A..96F}. Photometric conditions and seeing of $<$0.8 arcsec were required before execution of the observations could begin. 
 Due to the relatively high spectral resolution of XShooter ($\sim$35 km\,s$^{-1}$), host galaxy features were easily identified and removed in the reduction process. For one SN in our sample, SN 2013ct, due to the potential detection of a feature consistent with \ha\ emission (see Section \ref{SN2013ct} for more details), we also performed an additional extraction of the spectra using a custom-written pipeline. The two-dimensional data products from the REFLEX pipeline were used as input, but with sky-subtraction, extraction and flux-calibration performed order-by-order on the unresampled REFLEX images. This was to check carefully the host galaxy feature removal and relative flux calibration, and confirm that this was not causing the tentative emission feature that was seen. 

We scale the flux of our spectra to photometry taken close in time to the spectroscopic observations to obtain absolute flux-calibrated spectra. We obtained acquisition images using \textit{r'} and \textit{i'} band filters, similar to the filters used by the Sloan Digital Sky Survey \citep{2000AJ....120.1579Y}, and in the Johnson \textit{I} band filter. The SN magnitudes were determined using point spread function (PSF) photometry. The zero-points of the images were obtained using aperture photometry of comparison stars in the images, which were then calibrated by comparison to either catalogue magnitudes from the SDSS Data Release 10 \citep{2014ApJS..211...17A} for SNe 2012cg and  2012ht, or using the XShooter acquisition and guide camera zero-points for SNe 2012fr, 2013ct, 2013cs, 2013aa. XShooter zero-points were not available for \textit{r'}. Therefore, for the three SNe (2013ct, 2013cs, and 2013aa) that were observed in the \textit{r'} band, a transformation, using the relative flux throughputs of the SN spectra in the relevant filters, was applied to convert  \textit{r'} to Johnson \textit{R} for which an XShooter zero-point was available. 

The flux-calibrated spectra were corrected to rest-frame wavelengths using the heliocentric velocities given in Table \ref{tab:SN_info}. These values were obtained from the NASA/IPAC Extragalactic Database\footnote{http://nedwww.ipac.caltech.edu/} (NED) galaxy spectra, apart from SN 2012cg, which following \cite{2013MNRAS.436..222M} used the stellar velocity of \citet{2006AJ....131..747C} at the SN position. 

\begin{table}
  \caption{SN host galaxy information.}
 \label{tab:SN_info}
\begin{tabular}{@{}lccccccccccccccccccccccccccccc}
  \hline
  \hline
Name &Host galaxy&Vel.$_{helio}$$^a$ &Distance$^b$&E(\textit{B-V})$^c$\\
&&(\kms)&(Mpc)&(mag)\\
\hline
\hline
SN 2009ig&NGC 1015&2629$\pm$6&34.5$\pm$7.0&0.029\\
SN 2010gp&NGC 6240&7339$\pm$9&105.3$\pm$7.4&0.067\\
SN 2011ek&NGC  0918&1507$\pm$3&17.0$\pm$1.6&0.312\\
SN 2011iv&NGC 1404&1947$\pm$4&20.9$\pm$1.2&0.010\\
SN 2012cg&NGC 4424    &461$\pm$4&15.2$\pm$1.9&0.20$^d$\\
SN 2012cu&NGC 4772&1040$\pm$5&43.8$\pm$7.3&0.914$^d$\\
SN 2012fr&NGC 1365      &1636$\pm$1&19.1$\pm$0.9&0.018\\
SN 2012ht&NGC 3447a&1066$\pm$1&20.1$\pm$1.5&0.026\\
SN 2013aa&NGC 5643    &1199$\pm$2&18.1$\pm$3.6&0.169\\
SN 2013cs&ESO 576-017 &2771$\pm$1&44.1$\pm$3.1&0.082 \\
SN 2013ct&NGC 0428       &1152$\pm$2&12.1$\pm$1.2&0.025\\
\hline
\end{tabular}
 \begin{flushleft}
  $^a$Heliocentric velocities from NED. \\
 $^b$Distances calculated from cosmic microwave background redshifts from NED apart from redshift-independent distances for SN 2009ig \citep{1988ngc..book.....T}, SN 2011ek  \citep{2013AJ....146...86T}, SN 2011iv \citep{2013AJ....146...86T}, SN 2012cg \citep{2008ApJ...683...78C}, SN 2012cu \citep{2009AJ....138..323T}, SN 2012fr  \citep{2013AJ....146...86T}, SN 2013aa \citep{1988ngc..book.....T}, and SN 2013ct \citep{2013AJ....146...86T}. \\
 $^c$Galactic E(\textit{B--V}) values from \protect \cite{2011ApJ...737..103S}.\\
 $^d$Host galaxy extinction of E(\textit{B--V}) = 0.18 mag was found for SN 2012cg by \cite{2012ApJ...756L...7S}. For SN 2012cu, an additional host galaxy extinction of E(\textit{B--V}) = 0.89 mag was determined using the results of \cite{2015MNRAS.453.3300A} corrected to an R$_V$ of 3.1. The values quoted are the combined Galactic and host galaxy E(\textit{B--V}) values. \end{flushleft}
\end{table}

Following \cite{2007ApJ...670.1275L}, the optical regions of the XShooter spectra were rebinned to 3 \AA. The NIR portions of the spectra were rebinned to 6 \AA\ to increase the signal-to-noise ratio. A host-galaxy origin for any potential $\sim$1000 \kms\ \ha\ emission can be confidently ruled out since it would only contribute to the central spectral bin at the binned XSH dispersion.

\begin{figure*}
\includegraphics[width=13cm]{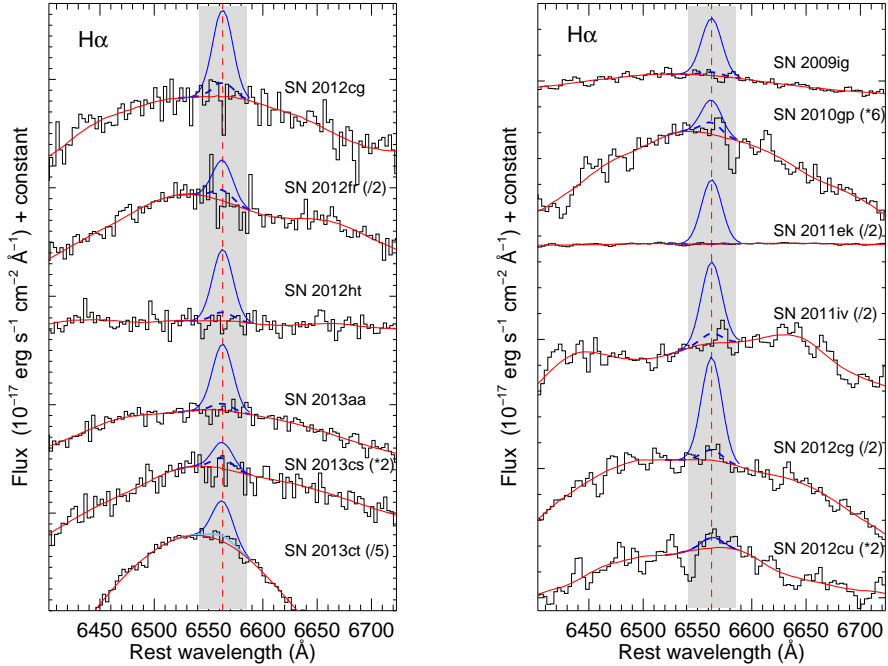} 
\caption{Left-hand panel: the H$\alpha$ spectral region of the SNe Ia with XShooter spectra at phases greater than +200 d. The black solid lines are the binned spectra (binned dispersion of 3 \AA) and the red solid lines are the calculated continuum fits as described in Section \ref{analysis}. The red vertical dashed line marks the rest position of H$\alpha$ (6563 \AA). The grey shaded region of $\sim$1000 \kms\ ($\pm$22 \AA) is where \ha\ emission is expected based on modelling efforts. The expected \ha\ emission from 0.05 \msun\ of material at +380 d as predicted by the modelling of \protect \cite{2005A&A...443..649M} is shown as a blue solid line for each SN and has been corrected for the distance and extinction towards each SN. The estimated 3$\sigma$ detection limit is shown for each SN as a blue dashed line. For presentation purposes, some spectra have been multiplied or divided by the constants specified in parentheses after the SN name. The detection of \ha\ for SN 2013ct is shown as a light blue solid line. Right-hand panel: the H$\alpha$ spectral region for the SNe Ia in our sample with FORS2 spectra (binned dispersion of 3.25 \AA) is shown. }
\label{spec_hsearch}
\end{figure*}

\subsection{FORS2 spectra}
For the FORS2 observations, a combination of grisms 300V and 300I (+OG590) and a 1 arcsec slit were used, yielding a resolution $R \sim 400$--$1000$ over a wavelength range of 3500--10500\,\AA. After bias subtraction and flat-fielding an optimal, variance-weighted extraction of the spectra was performed wite \textsc{iraf}\footnote{\textsc{iraf} is distributed by the National Optical Astronomy Observatory, which is operated by the Association of Universities for Research in Astronomy under cooperative agreement with the National Science Foundation.} task \textsc{apall}. The dispersion solution was established using arc-lamp exposures and cross-checked against night-sky emission lines. Spectrophotometric standard stars, usually observed during the same nights as the SNe, were used to perform a relative flux calibration of the spectra and to correct for telluric absorptions. 

To calibrate to an absolute flux scale and correct for slit losses, the spectra were scaled to match FORS2  $BVRI$ photometry of the SNe obtained close in time to the spectral observations (see Table~1). The SN magnitudes in the $BVRI$ frames were measured using PSF photometry, and calibrated using the FORS2 zero-points and extinction coefficients provided by the ESO data-quality monitoring\footnote{http://www.eso.org/observing/dfo/quality/FORS2/qc/zeropoints/zeropoints.html}.

As a result of the significantly lower spectral resolution of the FORS2 spectra ($\sim$450 km\,s$^{-1}$) compared to the XShooter spectra, the host-galaxy subtraction was more complex. For nine of the SNe Ia\footnote{SNe 2009le, 2010ev, 2010hg, 2010kg, 2011K, 2011ae, 2011at, 2011im, 2011jh} observed with FORS2, emission from a \Hii-region was visible in the two-dimensional spectra either at the SN location or very nearby. Therefore, a clean background subtraction could not be guaranteed for these objects and we remove these SNe from further discussion since the uncertain background subtraction could mask \ha\ emission from the SN itself. However, we note that these SNe Ia could have displayed narrow H features that went undetected. Therefore, the number of SNe Ia with narrow H features could be higher than measured in our sample. The six remaining SNe Ia observed with FORS2, and that have a clean background subtraction, are detailed in Table \ref{tab:spec_info}.  For these spectra, the dispersion was left at its native value of 3.25 \AA.

\section{Analysis}
\label{analysis}

We have searched for the presence of material that could have been removed from a non-degenerate companion star (via ablation and/or stripping) in a new late-time SN Ia spectral sample, as well as used model predictions to constrain quantitatively the presence or absence of solar abundance material.

\subsection{Searching for signatures of H}
\label{search_for_h}

We searched for the presence of narrow ($<$1000 \kms) lines of H (H$\alpha$, Pa$\alpha$, Pa$\beta$) in the spectra.  To do this an underlying continuum must first be defined. The underlying continuum of the spectra in these regions was fit using a second-order Savitzky-Golay smoothing polynomial \citep{1992nrfa.book.....P}, with the search wavelength region of $\pm$22 \AA\ of the H features excluded from the smoothing. This was to avoid biasing the continuum towards higher values if a H-emission feature was present. The smoothing scales used were significantly larger than the predicted widths of the narrow features, with widths of 80--140 \AA\ in the optical and $\sim$200 \AA\ in the NIR. Further analysis and the associated uncertainties on estimating the continuum for the \ha\ region is discussed in Section \ref{estimation}.

\begin{figure}
\includegraphics[width=7.5cm]{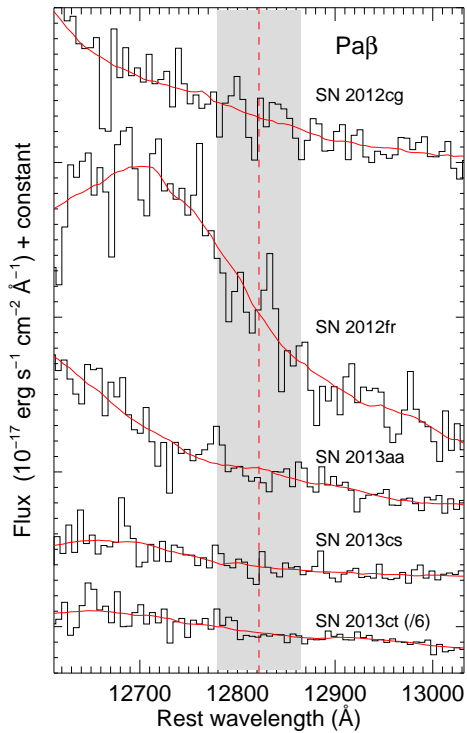} 
\caption{The Pa$\beta$ (12822 $\AA$) region is shown with the red vertical dashed line marking its rest position. The black solid lines are the binned spectra and the red solid lines are the smoothed continuum fits as described in Section \ref{analysis}.  The grey shaded region correspond to $\pm$1000 \kms of the Pa$\beta$ rest wavelength. No features consistent with narrow Pa$\beta$ emission are identified.}
\label{spec_hsearch_pb}
\end{figure}

The rebinned spectra and continuum fits at the position of \ha\ for our SN Ia sample are shown in Fig.~\ref{spec_hsearch}. 
The Pa$\beta$ (12822 $\AA$) regions of the XShooter spectra are shown in Fig.~\ref{spec_hsearch_pb}. The position of any potential Pa$\alpha$ (18750 $\AA$) feature falls in the middle of a telluric band and detection of SN flux is very difficult. Therefore, the only NIR H line we discuss is the Pa$\beta$ feature.   We did not detect any strong unambiguous \ha\ or Pa$\beta$ features in our sample when inspecting the appropriate wavelength regions. However, this does not exclude less prominent emission features being present. A quantitative estimate of the presence of \ha\ and associated non-detection limits is described in Section \ref{nondetect}.

We have estimated the potential effect of telluric features on detection of narrow features in the \ha\ and Pa$\beta$ spectral regions. The closest significant telluric feature (transmission $<$0.9) to the wavelength of \ha\ is a feature at $\sim$6519 \AA\, identified using the ESO SKYCALC Sky Model Calculator. For the SN in our sample with the lowest heliocentric redshift, SN 2012cg, this feature would lie at a rest frame wavelength of $\sim$6508 \AA. This is well outside the region of interest of $\pm$22 \AA\ of the \ha\ feature at 6563 \AA\ and the feature is weak enough not to be visible in our spectra. Therefore, we conclude that this has negligible effect on our results. We have also investigated potential telluric features in the Pa$\beta$ wavelength region and find no significant telluric features within the observed wavelength region.

\subsubsection{Model predictions of stripped/ablated material}
\label{mode_pre}
Based on the calculations of \cite{2000ApJS..128..615M},  \cite{2005A&A...443..649M} used a one-dimensional nebular-phase spectral code \citep[see][for more details]{1998ApJ...496..946K,2004A&A...428..555S,2005A&A...437..983K} to give a quantitative estimate of the \ha\ emission expected from different amounts of solar-abundance material stripped or ablated from a companion star. This solar abundance material was located in the inner 1000 \kms\ of the ejecta using the W7 explosion model \citep{1984ApJ...286..644N,1986A&A...158...17T}. The peak luminosity of the \ha\ line for 0.05 \msun\ of solar-abundance material was found to be $\sim$3.36 $\times$ 10$^{35}$ erg s$^{-1}$ \AA$^{-1}$ \citep{2005A&A...443..649M,2007ApJ...670.1275L} .

These model \ha\ luminosities were calculated at +380 d and the \ha\ emission is expected to be time-dependent \citep{2005A&A...443..649M}. However, \cite{2007ApJ...670.1275L} noted that since the optical depth to gamma rays should be higher at earlier times, the \ha\ emission should be stronger at epochs earlier than +380 d. Therefore, the estimated \ha\ flux at +380 d can be considered a lower limit for earlier spectra as long as the condition that the ejecta are transparent is fulfilled ($\gtrsim$200 d) and we can observe this low-velocity material. Sources of additional uncertainty in the modelling are discussed in Section \ref{model_uncertain}. 

For three SNe Ia in our sample (SNe 2009ig, 2011ek, and 2012ht), the spectra were obtained at epochs greater than +380 d (between +405 and +433 d). Given the small difference in the epoch studied in the model and the observed data, the model fluxes are assumed to be still applicable at these slightly later phases.

\begin{table*}
  \caption{Measured and derived quantities for \ha\ non-detection limits}
 \label{tab:derived_quant}
\begin{tabular}{@{}lccccccccccccccccccccccccccccc}
  \hline
  \hline
SN &Phase&\ha\ flux (0.05 \msun)$^a$&\ha\ flux limit (3$\sigma$)$^b$& Mass limit (3$\sigma$)$^c$  \\
&(d)&(10$^{-17}$ erg s$^{-1}$ cm$^{-2}$)&(10$^{-17}$ erg s$^{-1}$ cm$^{-2}$)&(\msun)\\
\hline
\hline
SN2009ig&405 &5.18$\pm$1.60&0.33&0.003\\
SN2010gp&277& 0.51$\pm$0.07&0.18&0.018 \\ 
SN2011ek&421&11.59$\pm$1.91&0.18&0.001 \\
SN2011iv&318 &14.70$\pm$1.55&1.87&0.006\\
SN2012cg&339 &18.57$\pm$3.90&3.09&0.008  \\
SN2012cg&343 & 18.57$\pm$3.90&1.98&0.005\\
SN2012cu&344&0.48$\pm$0.13&0.56&0.058\\
SN2012fr&357 &17.30$\pm$1.52 &4.79&0.013 \\
SN2012ht&433 &15.36$\pm$2.06&2.05&0.007  \\
SN2013aa&360&14.19$\pm$4.32&1.46&0.005 \\
SN2013cs&303 & 2.83$\pm$0.36&1.18&0.021 \\
SN2013ct&229& 42.46$\pm$7.32&5.66$\pm$0.98$^*$&   0.007$\pm$0.001$^*$ \\
\hline
\end{tabular}
 \begin{flushleft}
$^a$Expected integrated line flux of H$\alpha$ (with FWHM of 1000 \kms) from 0.05 \msun\ of solar metallicity material according to the models of \protect \cite{2005A&A...443..649M}.   \\
$^b$Estimated 3$\sigma$ upper limit on the integrated flux of H$\alpha$ (with FWHM of 1000 \kms) estimated as described in Section \ref{nondetect}.\\
 $^c$Derived 3$\sigma$ upper limits to the amount of solar metallicity material that could remain undetected. \\
 $^*$For SN 2013ct, the values quoted are the derived flux and mass for the detected feature at the position of \ha\ (with best-fitting FWHM of 850 \kms).  \\
\end{flushleft}
\end{table*}

We have calculated the expected \ha\ emission for this luminosity for each SN in our sample, using the distance and Galactic extinction values [corrected to the wavelength of \ha\ using the \cite{1999PASP..111...63F} extinction curve with R$_V = 3.1$] in Table \ref{tab:SN_info}. Host galaxy extinction corrections were not applied for most SNe Ia in our sample since we found no evidence for significant host galaxy extinction\footnote{Based on the analysis of \cite{2012ApJ...756L...7S}, we used a combined Galactic and host galaxy extinction of E(\textit{B-V}) = 0.20 mag for SN 2012cg. For SN 2012cu, a combined extinction of E(\textit{B-V}) = 0.914 mag was  used \citep{2015MNRAS.453.3300A}.}. This was based on an analysis of their maximum-light spectra, absolute magnitudes at peak, and locations within their host galaxy. 

 The expected integrated \ha\ fluxes for 0.05 \msun\ of H-rich material, assuming a Gaussian profile with a full width at half-maximum (FWHM) of 1000 \kms, and a central wavelength of 6563 \AA\ are given in Table \ref{tab:derived_quant}. The chosen value of the FWHM is consistent with the previous studies, as well as the model constraint of \cite{2005A&A...443..649M} that the H-rich material is located at velocities below $\sim$1000 \kms.  The expected \ha\ line profiles for 0.05 \msun\ of solar abundance material are plotted in Fig.~\ref{spec_hsearch} for each SN in our sample.

 \subsubsection{Estimation of contributions from narrow \ha}
 \label{estimation}
 
To estimate the amount of \ha\ that could be present, as well as set limits on any non-detections, we made model spectra combining the continuum fit with a Gaussian to represent a potential \ha\ emission feature. The Gaussian was set to have a fixed FWHM of 1000 \kms\ and fixed central wavelength of 6563 \AA, but a varying peak flux. A likelihood function was calculated as a function of varying peak flux in steps of 0.001 \msun\ within the range corresponding to $\pm$0.05 \msun\ (negative values are included to sample a complete peak-flux distribution). \cite{2005A&A...443..649M} and \cite{2013MNRAS.435..329L} calculated the expected flux at the position of \ha\ for differing input masses of stripped/ablated H-rich companion star material (0.01, 0.05, 0.1, 0.5 \msun), and found a linear relation between \ha\ flux and mass. Therefore, we use a linear scaling between \ha\ flux and mass in our analysis.

The largest uncertainty in the likelihood calculation is the estimation of the continuum fit. Therefore, to estimate the associated uncertainty we have calculated a number of continuum fits using the Savitzky-Golay smoothing polynomial with widths varying from 80 to 140 \AA. This range was chosen so has to be significantly larger than the width of the feature we are searching for  ($\pm$22 \AA) but not so big that the underlying continuum features are not well fit. A value of 100 \AA\ was used in the analysis of \cite{2007ApJ...670.1275L}.  The mean of these continuum fits is then used as input to the likelihood calculation.
 
The sigma entering the likelihood was calculated from the rms scatter per wavelength bin from the continuum fits (an estimation of the uncertainty in the continuum definition) combined with the scatter around the normalized continuum (an estimation of the spectral noise).

\subsubsection{Tentative detection of \ha\ in SN 2013ct}
\label{SN2013ct}

For one SN in our sample, SN 2013ct, we have identified a weak but broad feature at the position of \ha, consistent with the stripping/ablation model predictions of \ha\ emission with velocities of $\sim$600--1000 \kms. By varying the width of the feature, we found the strongest significance feature for an FWHM of 850 \kms. Using this FWHM and the method detailed in Section \ref{estimation}, we found that the \ha\ flux of SN 2013ct corresponded to $\sim$0.005 \msun\ of stripped or ablated H-rich material. 

Due to the strong underlying continuum for SN 2013ct and the potential detection of a narrow \ha\ feature, we investigated further fits to the underlying continuum. We expanded the continuum fitting models to include a Gaussian fit to the underlying continuum. The parameters of the Gaussian were set to be variable. The region of the spectrum included in the Gaussian was varied between $\pm$70 and $\pm$130 \AA\ of the centre of the underlying profile (6540 \AA). The region $\pm$1000 \kms\ from the rest wavelength of \ha\ was again excluded from the continuum fitting.

\begin{figure}
\includegraphics[width=8.2cm]{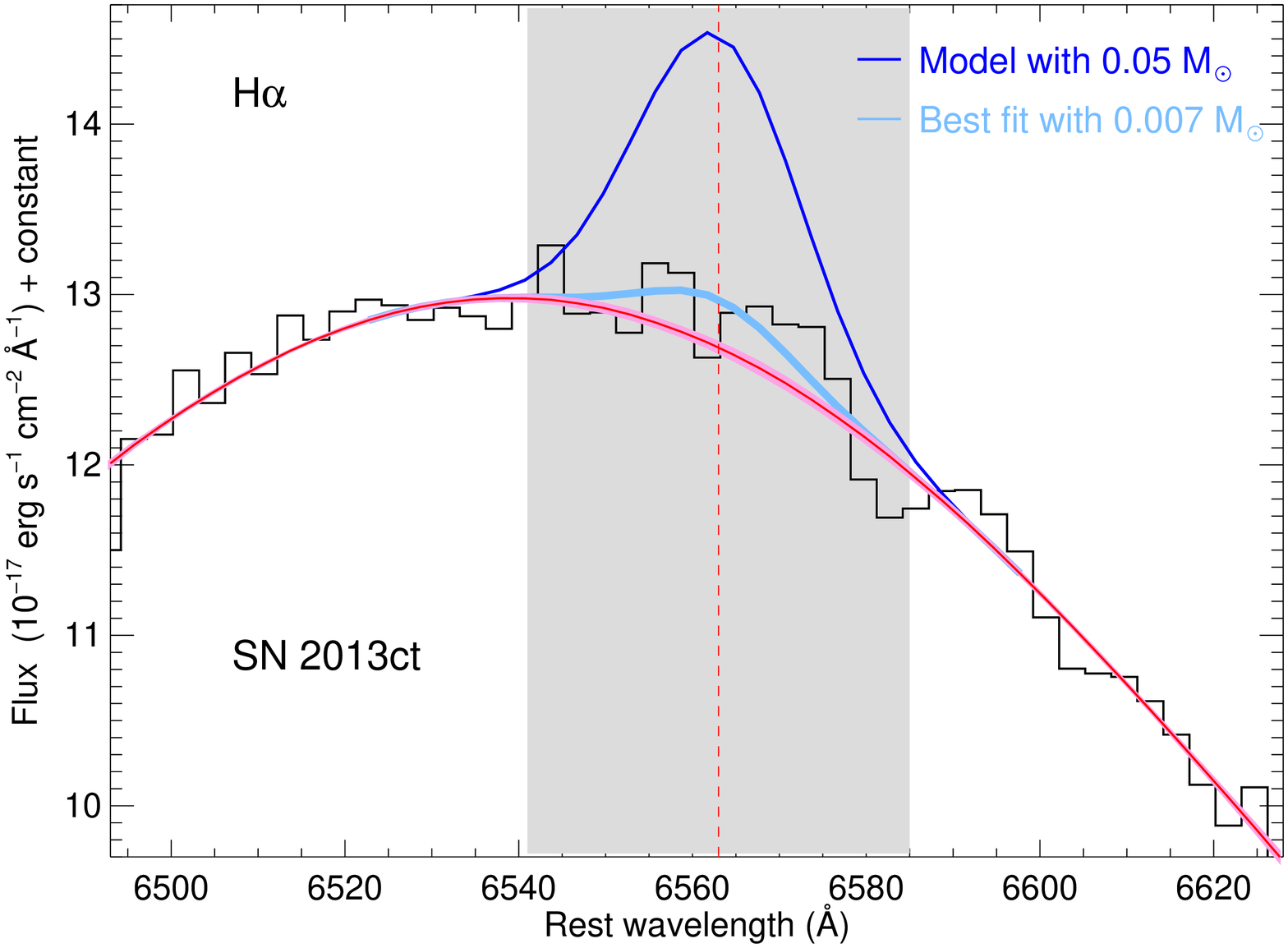}
\caption{The \ha\ region of the SN 2013ct spectrum at 229 d past maximum. The black solid line is the binned spectrum with the mean continuum fit over plotted as a red solid line. The pink shaded region is the standard deviation of the continuum fits. The red vertical dashed line marks the rest position of \ha\ (6563 \AA). The grey shaded region of 1000 \kms\ ($\pm$22 \AA) is where \ha\ emission is expected based on modelling efforts. The expected \ha\ emission from 0.05 \msun\ of material at +380 d as predicted by the modelling of Mattila et al. (2005) is shown as a blue solid line. The light blue solid line is the best-fitting detection of $\sim$0.007 \msun\ of H with an FWHM of 850 \kms. }
\label{13ct_plot}
\end{figure}

The mean continuum of the combined Savitzky--Golay and Gaussian fitting was used as the continuum in the likelihood calculation. The best-fitting for the \ha\ feature for SN 2013ct using this expanded continuum fitting is shown in Fig.~\ref{13ct_plot}. This \ha\ flux corresponds to $\sim$0.007$\pm$0.001 \msun\ of stripped or ablated H-rich material at a 3.7$\sigma$ significance.

To help determine the robustness of our \ha\ detection for SN 2013ct, we performed two additional tests. First, we searched for similar features at different rest wavelengths by varying the wavelength of the expected feature within $\pm$120 \AA\ of the detected \ha\ position. We used the same analysis method as for the search at the position of \ha, including excluding the search region from the continuum fitting. Secondly, we tested if a feature of a similar significance was detected on top of another broad nebular emission feature (the 5800 \AA\ feature) in the SN 2013ct spectrum. This was to determine if residuals in the fitting of the continuum for the broad emission features would result in a similar feature to that seen at the position of  \ha.  In both cases, no significant detection ($\geq$3$\sigma$) is associated within any feature outside of the \ha\ search region. However, we caution that the significance of the \ha\ detection is under the assumptions of the fitting to an unknown underlying continuum, as well as the predictions of the models of the position and velocity distribution of H-rich material. 

If our measurements of the SNe are assumed to be independent, then as the sample size increases, there is an increased probability of finding a 3$\sigma$ detection by chance. We used the correction of \cite{sidak_1967} to estimate the increased significance needed to claim a 3$\sigma$ detection for one object out of the eight SNe Ia in our combined new data and literature sample for which the mass detection limit at the position of \ha\ is $\leq$0.007 \msun\ (equivalent to the SN 2013ct detection). We found that a 3.7$\sigma$ detection in any one spectrum is equivalent to a 3.1$\sigma$ detection if we make eight comparisons, which we use as our significance value for this detection.

\begin{figure*}
\includegraphics[width=17.cm]{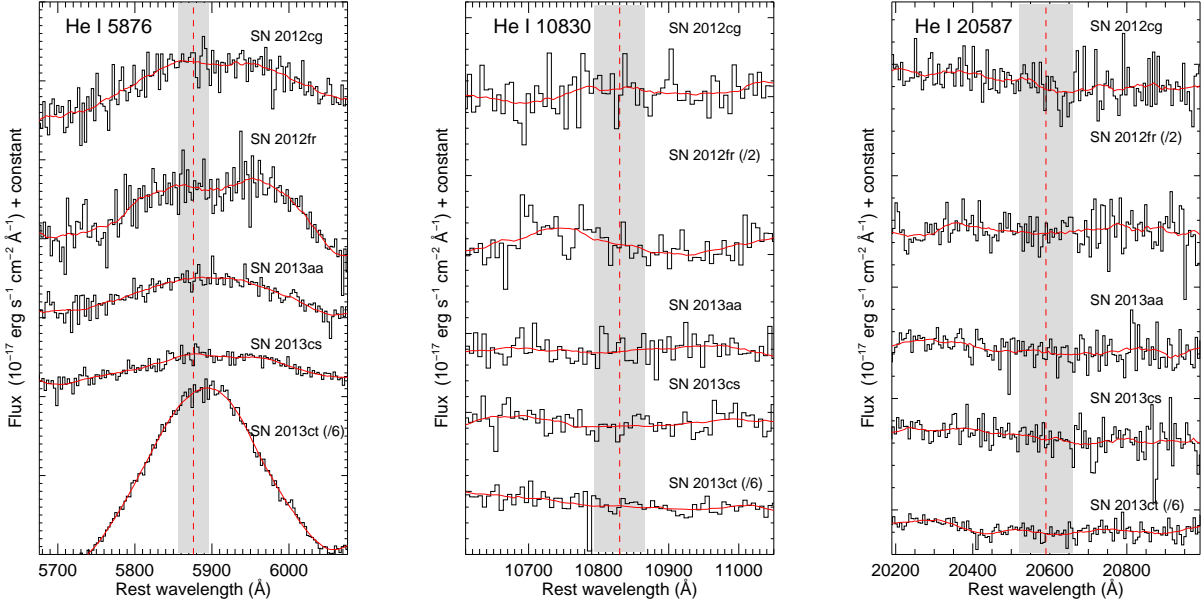}
\caption{The \Hei\ spectral regions for five SNe Ia in our sample at phases greater than +220 d is shown. The black solid lines are the binned spectra and the red solid lines are the smoothed continuum fits as described in Section \ref{analysis}.  The red vertical dashed lines mark the rest position of \Hei\ lines and the grey shaded regions correspond to $\pm$1000 \kms.  No \Hei\ emission is identified in our spectra. We note that the \Hei\ 20590 \AA\ region is moderately affected by telluric absorption features and therefore, our results are based predominantly on the \Hei\ 5876 and \Hei\ 10830 \AA\ features. }
\label{spec_hesearch}
\end{figure*}

\subsubsection{\ha\ detection limits}
\label{nondetect}
The rest of the SNe Ia in our sample showed no obvious \ha\ emission. Using the same method as for SN 2013ct, we determined that the Gaussian fits at the position of \ha\ for the rest of the sample are consistent with zero within the 3$\sigma$ uncertainties. We placed 3$\sigma$ upper limits on the strength of features that could have remained undetected at the position of \ha. 

The line profiles representing these flux limits are shown in Fig.~\ref{spec_hsearch}, and values for the integrated \ha\ flux and corresponding mass limits are given in Table \ref{tab:derived_quant}. The 3$\sigma$ limiting mass range for the SN sample is 0.001--0.058 \msun. The bright, nearby SN 2012cg was observed with both XShooter and FORS2 and we obtain a consistent mass limit ($<$0.010 \msun) for both spectra.

Since these limits are strictly statistical, \cite{2015A&A...577A..39L} estimated how systematic uncertainties from our lack of knowledge of the underlying spectral continuum could affect the measured \textit{F}(3$\sigma$) values of \ha\ emission for SNe 2011fe and 2014J. They suggested, using an inspection by eye, that the limits on the masses of stripped/ablated material could be 2--3 greater than their measured values.  

As discussed in Section \ref{search_for_h}, to minimize the uncertainty in the selection of the continuum in our sample, we have excluded the region of $\pm$22 \AA\ with respect to \ha\ when smoothing the spectra to avoid biasing the continuum fit of the search wavelength region. We have also tested our continuum fits by checking for similar strength features outside the \ha\ search region, as well as on top of the broad 5800 \AA\ nebular feature, and no detection was made. Therefore, we have chosen to use our original \textit{F}(3$\sigma$) values calculated from the probability distributions to measure \textit{M}(3$\sigma$) and these are the values we quote in Table \ref{tab:derived_quant}.

\subsection{Searching for signatures of He}
The companion star to the exploding white dwarf could also potentially be a He-rich star instead of a H-rich star, which could result in the presence of narrow ($<$1000 \kms) He lines in late-time SN Ia spectra. Less material is expected to become unbound than in the H-rich case; just 0.02--0.06 \msun\ is predicted by the models \citep{2012ApJ...750..151P,2013ApJ...774...37L}.

To investigate the presence of He emission at late times, we focus first on the \Hei\ 5876, 10830, and 20587 \AA\ features in the XShooter VIS and NIR spectra. The wavelength regions for the \Hei\ lines are shown in Fig.~\ref{spec_hesearch}. The \Hei\ 5876 and 10830 \AA\ spectral regions are found to be relatively free of telluric absorption features. However, the \Hei\ 20590 \AA\ region is moderately affected. For completeness, we show this region in Fig.~\ref{spec_hesearch} but do not base our results on this region alone. We did not have model predictions of the peak luminosity for these He features but from Fig.~\ref{spec_hesearch} it is clear that there are no strong \Hei\ signatures detected in our SN Ia sample within the velocity range predicted by \cite{2012ApJ...750..151P}. 

\cite{2013MNRAS.435..329L} detailed how \Oi\ and \CaiiF\ emission features are promising probes of stripped/ablated He-rich material.  We have also searched for \Oi\ 6300, 6364 \AA\ and \CaiiF\ 7291, 7324 \AA\ emission features with velocities of $\pm$1000 \kms\ of the rest wavelength in our XShooter and FORS2 spectra but did not find any features consistent with the qualitative model predictions. However, further analysis, when robust spectral modelling predictions of peak luminosities are available, will be necessary to put flux (and mass) limits on these He, Ca and O non-detections.

\section{Discussion}
\label{discussion}
Using the expected \ha\ luminosities from a time-dependent nebular synthesis calculation for solar abundance material confined within 1000 \kms\ \citep{2005A&A...443..649M}, we have quantified the presence or absence of H-rich companion star material in 11 SNe Ia for which late-time spectra were obtained. For 10 SNe Ia in the sample, we did not find evidence of H-rich companion star material in their late-time spectra. Under the assumptions of the modelling, we can place limits of $<$0.001--0.058 \msun\ on the mass of H-rich material that could remain undetected. For one SN in our sample, SN 2013ct, we have made a tentative detection (3.1$\sigma$) of emission at the position of \ha\, corresponding to 0.007$\pm$0.001 \msun\ of H-rich material. Combining these new data with previous samples \citep{2005A&A...443..649M,2007ApJ...670.1275L,2013MNRAS.435..329L,2015A&A...577A..39L,2013ApJ...762L...5S}, this means one  potential detection of \ha\ emission out of 18 `normal' SNe Ia that have been studied at late times to look for signatures of H-rich material. 

Emission at the wavelength of Pa$\beta$ was also investigated for five SNe Ia in the sample with suitable NIR spectral coverage and no clear emission was identified. However, as noted in  \cite{2013MNRAS.435..329L}, the mass limits obtained from Pa$\beta$ are less constraining than for \ha.

We do not detect strong signatures of \Hei\ 5876, 10830, 20587 \AA\ features within a velocity range of $\sim$1000 \kms. We also do not detect \Oi\ or \CaiiF\ emission features, which were suggested by \cite{2013MNRAS.435..329L} to be promising tracers of He-rich material.  \cite{2015A&A...577A..39L} estimated very rough limits on the He mass present for SNe 2011fe and 2014J by adjusting the H mass limits for the He to H number density ratio assuming solar metallicity. For the tightest mass limits for SN 2011fe in \cite{2015A&A...577A..39L}, all He-rich companion stars from the models of \cite{2012ApJ...750..151P} 
and \cite{2012A&A...548A...2L,2013ApJ...774...37L} can be ruled out based on the non-detection of \Oi\ emission. However, since the He-mass predictions obtained are simple extrapolations from models of H-rich solar abundance material placed in the inner $\sim$1000 \kms\ of the SN ejecta and the He-rich material is expected to be at higher velocity than this, they should be taken as very rough upper limits. 

\subsection{First detection of \ha\ in the late-time spectrum of a normal SN Ia?}
\label{SN2013ct_discussion}
The detection of a potential \ha\ feature consistent in velocity ($<$1000 \kms) with the models of \cite{2012ApJ...750..151P} and \cite{2012A&A...548A...2L} is suggested for SN 2013ct. Using the models of \cite{2005A&A...443..649M}, this is estimated to be equivalent to 0.007$\pm$0.001 \msun\ of solar abundance material being present at velocities below 1000 \kms. However, the models of  \cite{2012ApJ...750..151P} and \cite{2012A&A...548A...2L} have shown that if the companion star to the exploding white dwarf was an MS star in any realistic binary scenario, \textit{at least} 0.1--0.2 \msun\ of material should be removed from the companion star after explosion. The companion star can not be artificially moved to greater separations to reduce the amount of stripped/ablated material because then the necessary condition of RLOF would not be fulfilled. For an RG companion, $\sim$0.6 \msun\ of material is predicted to be stripped/ablated from the companion in RLOF. For systems transferring mass via a stellar wind, $>$0.5 \msun\ of material  would be removed from the companion star \citep{2012ApJ...750..151P}

Therefore, in order to explain an emission feature corresponding to $\sim$0.007 \msun\ of H-rich material, an additional $\sim$0.1--0.6 \msun\ of material would have to be present but not visible as a narrow \ha\ emission within 1000 \kms. A tail of material extending to higher velocities could hide some material that would not contribute significantly to the observed narrow \ha\ feature. However, the models predict that the amount of stripped or ablated material present in a high-velocity tail is expected to be low  \citep{2012ApJ...750..151P,2012A&A...548A...2L}. 

Another possibility is that the majority of the H-rich companion star material is not sufficiently powered by radioactive heating to produce \ha\ emission. This may be because there is little SN ejecta remaining below 1000 \kms\ at these times \citep{2012A&A...548A...2L}. Therefore, while the presence of a narrow \ha\ feature suggests a H-rich non-degenerate companion star to the white dwarf for SN 2013ct, further constraints on the companion star properties can not currently be placed. The uncertainties of the model and observations are detailed further in Section \ref{model_uncertain}.

\subsection{Are H-rich SD systems sub-dominant for producing normal SNe Ia?}
\label{SD_models}

Unfortunately, little early-time data are available for SN 2013ct to determine its detailed properties. The SN was discovered on 2013 May 10 but not spectroscopically classified until 2013 May 22, where the best near-infrared spectroscopic match was a normal SN Ia approximately 20 d after maximum light \citep{2013ATel.5081....1M}. Therefore, while this spectrum appears similar to other normal SNe Ia, subtle differences that may have been apparent with a larger data set are not quantifiable.

Although the first tentative detection of \ha\ emission in a late-time spectrum of a `normal' SN Ia is very interesting, the absence of \ha\ emission in the late-time spectra of 17 other SNe Ia is also worthy of discussion. Using the models of \cite{2005A&A...443..649M}, stripped/ablated mass limits in the range 0.001--0.058 \msun\ were determined for these 17 SNe Ia.  If we assume that the stripping/ablation and spectral model calculations are correct, then these limits are sufficient to rule out all MS and RG companions transferring mass via RLOF or a stellar wind at all plausible separations.

However, the  \ha\ detection equivalent to $\sim$0.007 \msun\ of H-rich material for SN 2013ct potentially weakens these constraints, since it suggests that $\sim$0.1--0.6 \msun\ of H-rich companion star material was hidden for this SN. Therefore, H-rich material may have also been present in the other 17 SNe Ia but also hidden.

The spin-up/spin-down scenario of  \cite{2011ApJ...730L..34J} and \cite{2011ApJ...738L...1D} could reduce the amount of H-rich material stripped from a non-degenerate companion star at the time of explosion. This involves a white dwarf becoming spun-up by the mass accreted from its companion star, resulting in a stable white dwarf above the Chandrasekhar mass. This gives time for the companion star to evolve and contract before the SN explosion. Therefore, at the time of explosion, the companion star is much smaller and more tightly bound, significantly reducing the amount of material removed by the impact of the SN ejecta. However, some fine-tuning is necessary to produce this scenario and as the sample of SNe Ia without H emission grows, it is becoming increasingly unlikely that this can explain all the observed non-detections.

It has also been suggested that the broad underlying emission feature at the position of \ha\ in our sample (usually attributed to \FeiiF\ 6559 \AA\ emission) may have a contribution from a broad \ha\ emission component \citep{2015MNRAS.450.2631M}. However, in this case, the H-rich material would have to be present at much higher velocities than those currently predicted by modelling efforts  \citep{2012A&A...548A...2L,2012ApJ...750..151P}. Further studies are necessary to determine if this is physically plausible. 

However, it is likely that the sub-class of SNe Ia, SNe Ia-CSM, do result from an SD progenitor channel. The most likely scenario is that of a symbiotic system involving an RG or asymptotic giant branch star \citep[][]{2003Natur.424..651H,2012Sci...337..942D}. This scenario would produce the necessary CSM to explain the interaction features seen in their spectra such as broad H emission extending till late times \citep{2013ApJ...772..125S}. If SNe Ia-CSM and at least some so-called `normal' SNe Ia originate from a SD scenario, then it may not be surprising to identify weaker H features, such as that detected for SN 2013ct,  in some SNe Ia (coming from stripped/ablated companion material instead of pre-explosion mass loss).

\subsection{He-rich companion stars}
\label{He_discuss}

Another solution for producing `normal' SNe Ia through the SD channel and avoiding H contamination is to invoke a He- instead of H-rich companion star. The latest simulations of the interaction between a He-rich companion star and the SN ejecta predict stripped/ablated masses of 0.02--0.06 \msun\ \citep{2012ApJ...750..151P,2013ApJ...774...37L}. \cite{2015A&A...577A..39L} put limits on the mass of He-rich material present in SNe 2011fe and 2014J of $<$0.002 and $<$0.005 \msun, respectively. Under the assumption of RLOF, these limits rule out the He-rich companion star models of \cite{2012ApJ...750..151P} and \cite{2013ApJ...774...37L} for both SNe Ia.

We have presented a qualitative discussion of the non-detection of He emission in our late-time spectral sample of 11 SNe Ia -- we do not find obvious features that could be attributed to \Hei, \CaiiF\ or \Oi\ emission. However, as cautioned in \cite{2015A&A...577A..39L}, these limits are based on a rough correction from H- to He-rich material, and have not been modelled explicitly. Therefore, we await more detailed spectral modelling to place limits on the presence of He-rich material  from a companion star swept-up in the SN ejecta.

Companion stars with He-rich outer layers are also present in the `double-detonation' scenario, where a thin layer of He on the surface of the primary white dwarf is responsible for the first detonation \citep{1982ApJ...253..798N,2014ApJ...785...61S}. This He material is accreted from a He star, He WD, a CO white dwarf with a thin layer of He on its surface, or it may be already present on the primary white dwarf surface. At the time of the subsequent detonation of the core,  the He on the surface is expected to have velocities of $>$20000 \kms. Therefore, this material is not expected to be visible at the low velocities studied here, and the non-detection of He features in the late-time spectra does not place constraints on the presence of high-velocity He-rich material in the `double-detonation' scenario. Detection of this high-velocity He is also unlikely in early-time observations because of insufficient heating of this material so far out in the ejecta to cause He emission lines to be observed.

\subsection{Sample selection}
The SNe Ia in our late-time spectral sample were selected for observation based on their proximity ($z\lesssim 0.025$) and visibility at Paranal at $>$200 d after maximum light.  If an SN Ia was significantly sub-luminous at maximum light then it would not have been scheduled for observations because it would be deemed too faint at $>$200 d after maximum light. This would bias our sample towards more luminous SNe Ia. However, the SNe Ia in our sample span a wide range of host galaxy types from early- to late-type galaxies, suggesting a spread in SN luminosity \citep{1995AJ....109....1H,1996AJ....112.2391H,2000AJ....120.1479H,1999AJ....117..707R,2006ApJ...648..868S}. SN 2011iv which occurred in an elliptical galaxy was a low-luminosity SN Ia \citep{2012ApJ...753L...5F}. Therefore, we conclude that our sample is not significantly biased towards more luminous events, and covers a range of `normal' SN Ia luminosities.  There is also no reason to expect that less-luminous SNe Ia are more likely to have H or He features present in their spectra. In fact, recent work suggested the opposite -- that it is likely that more luminous SNe Ia occur preferentially through SD channels \citep{2013ApJS..207....3S,2014MNRAS.444.3258M,2015A&A...574A..61L}. 

\subsection{Observational and model uncertainties}
\label{model_uncertain}
In this section, we discuss the uncertainties and limitations of the observations and modelling that could cause the results and interpretation of our analysis to be less constraining. A number of independent simulations have been made of the amount of stripped/ablated material, and its velocity distribution, that is expected to be removed from a non-degenerate companion star after explosion \citep{2000ApJS..128..615M,2007PASJ...59..835M,2008A&A...489..943P,2010ApJ...715...78P,2012ApJ...750..151P,2012A&A...548A...2L,2013ApJ...774...37L}. There is now reasonable agreement among the different groups for these masses and velocity distributions for different companion star setups. However, only one analysis has been done to determine the spectral line strengths associated with different mass and velocity values \citep{2005A&A...443..649M,2013MNRAS.435..329L}. This study was based on the earliest impact simulations of \cite{2000ApJS..128..615M}. However, the amount and velocity distributions of the stripped/ablated material have not been updated drastically in more recent modelling, and this should not significantly affect the spectral modelling results.

The largest uncertainty in the estimation of the presence of H- or He-emission features in the observed spectra is the calculation of the underlying continuum. We have used a second-degree Savitzky--Golay smoothing polynomial \citep{1992nrfa.book.....P} to fit the underlying continuum, as was used in previous late-time narrow \ha\ searches. This gives a good fit to the underlying continua in our sample, even in the case of broad underlying emission features. For SN 2013ct, we expanded our analysis of the continuum fitting to include also broad Gaussian fits to the underlying spectral feature, and again found a significant detection. However, given the tentative nature of the detection of \ha\ emission in SN 2013ct, we could not completely exclude the possibility that the continuum fits results in a residual consistent with the detected \ha\ emission. To determine how likely this is, we performed tests, detailed in Section \ref{SN2013ct}, looking for similar strength features at different wavelengths around \ha\ and also on top of a different broad emission feature at $\sim$5800 \AA. In neither case, was a similar feature found. 

With regard to the spectral synthesis models, the addition of the H-rich material potentially swept-up from a non-degenerate companion star in the models of \cite{2005A&A...443..649M} and \cite{2013MNRAS.435..329L} is somewhat ad hoc -- it is made by adding varying amounts of solar metallicity material with velocities $<$1000 \kms\ in a W7 density model, artificially increasing the density in the innermost region. It is uncertain if this is consistent with the density structure obtained from the three-dimensional modelling of the impact of the SN ejecta on the companion star. 

In particular, a major source of uncertainty in the spectral synthesis modelling is whether the low-velocity H-rich material is sufficiently powered by radioactive heating to produce H$\alpha$ emission. This depends on the location of the H-rich material relative to the radioactive material; if they are not co-located then H-rich companion star material may be present in the ejecta but not observable.  \cite{2005A&A...443..649M} performed the spectral line strength calculations at an epoch of +380 d. In the range of 150--300 d, the SN envelope is expected to become transparent to gamma-rays and enter a positron-dominated phase \citep{2009MNRAS.400..531S}. However, this is not expected to be the case for the central high-density regions, where the H is located. In these high-density regions, the optical depth to gamma-rays is likely to be high enough to power the H lines \citep{2005A&A...443..649M}, but more detailed modelling need to be carried out to confirm this. Indeed, the tentative detection of low-velocity H emission in SN 2013ct, corresponding to $\sim$0.007 \msun\ of H-rich material, suggests that $>$0.1--0.6 \msun\ of stripped/ablated material must be present but not observable.

The one-dimensional models of \cite{2005A&A...443..649M}  and \cite{2013MNRAS.435..329L} also assume spherical symmetry. However, the impact simulations show that the stripping and ablation of material from the companion star is not symmetric; the material is predominantly confined to the downstream region behind the companion star \citep{2000ApJS..128..615M,2008A&A...489..943P,2012A&A...548A...2L,2013ApJ...774...37L,2012ApJ...750..151P}. Since the ejecta are assumed to be optically thin at the late phases studied here ($\gtrsim$200 d), the detection of the swept-up material is not viewing angle dependent. However, the predicted asymmetry of the unbound material could affect the shape and wavelength of the observed line profiles. As discussed in \cite{2015A&A...577A..39L}, the model used by \cite{2005A&A...443..649M} and \cite{2013MNRAS.435..329L} also included only a limited number of elements, ionization states and atomic levels. No macroscopic mixing of the companion material was included in the spectral modelling either, which could affect the predicted shape and flux of the emission lines but is not expected to be a dominant source of uncertainty. 

Therefore, given the discussed uncertainties and limitations, future modelling, using the three-dimensional simulated ejecta structure of \cite{2012ApJ...750..151P} and \cite{2012A&A...548A...2L,2013ApJ...774...37L} as input to a multi-dimensional radiative transfer calculation for computing expected spectral fluxes, is of vital importance to confirm the model predictions, and hence observational mass limits. 

\section{Conclusions}
\label{conclusions}

We have presented a search for the presence of H- and He-rich material stripped or ablated from a non-degenerate companion star in new late-time spectra of 11 SNe Ia, obtained at the VLT+XShooter and the VLT+FORS2. The observed fluxes (or limits) at the position of \ha\ have been converted to masses using the spectral synthesis modelling described in \cite{2005A&A...443..649M} and \cite{2013MNRAS.435..329L}.  
Our main results are as follows.

\begin{enumerate}
\item We find evidence at the 3.1$\sigma$ level of \ha\ emission with a best-fitting FWHM of $\sim$850 \kms\ for one SN Ia in our sample, SN 2013ct. This corresponds to 0.007$\pm$0.001 \msun\ of H-rich material stripped/ablated from a non-degenerate companion star.
\item This mass is much lower than expected for MS+WD or RG+WD progenitor systems, suggesting at least 0.1 \msun\ of H-rich material is present in SN 2013ct but not observed as narrow \ha\ emission.
\item We find no evidence of H emission (\ha, Pa$\beta$) in the late-time spectra of 10 other SNe Ia, bringing the total sample with no H emission detected to 17 SNe Ia.
\item Upper limits on the stripped/ablated mass of solar abundance material of 0.001--0.058 \msun\ are placed for these SNe Ia.
\item These upper mass limits of H-rich solar abundance material are inconsistent with MS or RG companion stars transferring mass via RLOF or wind-driven accretion (under the assumptions of current modelling).
\item No signatures of He-rich material in the form of \Hei, \Oii\ or \CaiiF\ emission lines are identified. However, spectral modelling of the expected flux of lines from He-rich material is not available.
\end{enumerate}

While future observational studies will increase the sample size of SNe Ia with the necessary late-time observations, and perhaps identify narrow \ha\ emission in more objects, major future improvements are also likely to come from the next generation of spectral-synthesis modelling, allowing us to confirm (or adjust) these mass limits and determine if the SD scenario for producing the majority of SNe Ia is really in jeopardy. 

\section{Acknowledgements}
KM acknowledges the support of a Marie Curie Intra European Fellowship, within the 7th European Community Framework Programme (FP7). ST acknowledges the support of TRR 33 ``The Dark Universe'' of the German Research Foundation (DFG). MS acknowledges support from the Royal Society and EU/FP7-ERC grant no [615929]. Based on data taken at the European Organisation for Astronomical Research in the Southern hemisphere, Chile, under program IDs: 086.D-0747(A), 087.D-0161(A), 088.D-0184(A), 089.D-0821(A), 090.D-0045(A), 091.D-0600(A), 091.D-0764(A), and 092.D-0632(A). This research has made use of the NED which is operated by the Jet Propulsion Laboratory, California Institute of Technology, under contract with the National Aeronautics and Space Administration. 

\bibliographystyle{mn2e}
\bibliography{astro}

\end{document}